\documentclass[a4wide]{article} 
\usepackage[tracking=true]{microtype}
\usepackage[utf8]{inputenx}
\usepackage{mathptmx}
\usepackage[UKenglish]{babel}      
\usepackage{siunitx}               
\usepackage{url}                   
\usepackage{amsmath}               
\usepackage{amssymb}               
\usepackage{array}                 
\usepackage{mathtools}             
\usepackage{floatrow}              
\usepackage{graphicx}
\usepackage{subscript}             
\usepackage{booktabs}              
\usepackage{xcolor}                
\usepackage{multicol}              
\usepackage{multirow}              
\usepackage{subfigure}             
\usepackage{hyperref}              
\usepackage{amsthm}
\usepackage[%
english,%
algoruled,%
linesnumbered%
]{algorithm2e}                     

\newcommand{\norm}[1]{\left\|#1\right\|}
\newcommand{\abs}[1]{\left|#1\right|}

\newtheorem{definition}{Definition}

\begin{document}

\title{
  Highly Robust Clustering of \\
  GPS Driver Data for \\
  Energy Efficient \\
  Driving Style Modelling
  \thanks{
    The authors thankfully acknowledge the financial support of this work by
    the German Federal Ministry of Education and Research under grant
    no.~05M13ICC (RESY).}
}

\author{%
  Michael Breu{\ss} and Laurent Hoeltgen and \\
  Ali Sharifi Boroujerdi and
  Ashkan Mansouri Yarahmadi
}

\maketitle


\begin{abstract}
This paper presents a novel approach to distinguish driving styles with respect
to their energy efficiency. A distinct property of our method is that it relies
exclusively on Global Positioning System (GPS) logs of drivers. This setting is
highly relevant in practice as these data can easily be acquired.\par
Relying on positional data alone means that all derived features will be
correlated, so we strive to find a single quantity that allows us to perform the
driving style analysis. To this end we consider a robust variation of the so
called jerk of a movement. We show that our feature choice outperforms other
more commonly used jerk-based formulations and we discuss the handling of noisy,
inconsistent, and incomplete data as this is a notorious problem when dealing
with real-world GPS logs.\par
Our solving strategy relies on an agglomerative hierarchical clustering combined
with an L-term heuristic to determine the relevant number of clusters. It can
easily be implemented and performs fast, even on very large, real-world data
sets. Experiments show that our approach is robust against noise and able to
discern different driving styles.
\end{abstract}

\section{Introduction}
\label{hybb-intro}
The driving style has a significant impact on the fuel consumption of a car.
Intelligent hybrid cars could adapt to the driving style of the conductor to
maximise their mileage. These optimisations could be manifold and range from an
efficient assistance of the electric engine to suggestions on energy-optimal
routes~\cite{MSS2015,PB2015}. In this work we aim to provide a significant step
towards integrating the driving style as an additional constraint into this
objective. We analyse driving behaviour with respect to energy efficiency and
provide an automated way to classify it.\par{}
To allow a broad applicability of our methods, we rely on data that can easily
be obtained in any vehicle, as e.g.\ Global Positioning System (GPS) data logs.
Such setups are also attractive due to their relatively low cost and the already
abundant availability of GPS enabled devices. However, these benefits are at a
certain price. Insufficient accuracy yields noisy samples. Hardware failures may
result in a partial or total loss of the data. Furthermore, classifying the
driving style without environmental information is a quite challenging task.
Driving at a constant $\SI{120}{\kilo\metre\per\hour}$ on a large motorway over
lowland is surely an energy efficient way to travel but this may be doubtful
when the motorway goes over rolling hills. Traffic jams on motor ways lead to
stop and go traffic, which is visible in the logs by frequent small variations
in the acceleration and velocity, strongly resembling noise. However, the exact
cause of such variations cannot always be deduced with absolute certainty. The
driver may as well be looking for a free spot for parking, or his car may simply
have broken down. An accurate and robust model must be able to handle these
difficulties.\par{}
Fortunately, notable advances in modelling and understanding driver behaviour
have been made in recent years, yet mainly for use in the area of traffic safety
engineering. We refer to \cite{PL1999,QLP2012,QLR2010,RWGB2005,SBH2008} among
the vast amount of literature in that highly active field of research. Although
the objective in that field is completely different than in our case, let us
still review some works in more detail as we can identify at a technical level
some methods related to our approach.\par
In \cite{PL1999,SBH2008} the authors suggest the usage of sensors and simulators
to identify driver movements and to predict their behaviour in the forthcoming
seconds. These predictions can for example be used to prevent collisions. Both
works take a probabilistic approach by analysing dynamic and hidden Markov
models. The findings presented in \cite{IYOZ2005,RWGB2005} classify drivers
according to their imminent risk on the traffic. This classification is obtained
by comparing characteristic features of a given driver to information gathered
from other vehicles in the vicinity. While the authors of \cite{IYOZ2005} use
statistical measures, the authors of \cite{RWGB2005} combine a clustering
algorithm with complex neuroscale and Bayesian factor analyses. A similar
research is also performed in \cite{VVD2012}, where cars are analysed for
potentially dangerous behaviour. However, the focus of the latter work lies more
on the communication protocols between the traffic participants and not on their
classification. All these approaches have in common that they collect
significant amounts of data from a heterogeneous pool of sources and that they
are designed to analyse the conductor during driving. Many works in the past
have also tried to model the conductor himself by analysing his reactions in
various settings, see \cite{M1985,MNOW2007}. The authors of the latter work use
a car with specialised sensors to measure various information on the drivers.
These include position and velocity of the car as well as operation patterns of
the gas and brake pedal. The analysis is performed by means of Gaussian mixture
and optimal velocity models. In contrast to the first cited references these
latter works process the data only after collecting it.\par{}
Let us now turn to the use of GPS data. Modern potent tracking devices such as
navigation systems, cell phones or smartwatches offer huge amounts of
information that may be processed to distinguish driving styles. Surprisingly,
only few works consider this much simpler approach to restrict the study
exclusively on positional information provided by the tracking devices. In
\cite{LYC2010} the authors seek reckless taxi drivers by analysing the velocity
of their cabs and the regions that they pass. The authors of \cite{ZLZC2011}
have similar goals but limit themselves to analyse the routes taken. Here, a
taxi driver becomes suspicious when his route deviates strongly from those taken
by the majority of his colleagues. Let us remark that none of these scenarios
require real-time processing capabilities. The information can be stored inside
a database first and processed afterwards. In these settings missing additional
environmental data can, to a certain extent, be compensated by increasing the
amount of positional information. Furthermore, an offline processing of the data
allows to use more powerful hardware than available inside a car.\par{}
%
\paragraph{Our contribution}
%
To our best knowledge we propose in this work for the first time in the
literature a method for the analysis of energy efficient driving styles at hand
of GPS data collected by drivers. To this end we combine the two different
setups presented in the previous paragraphs. Similarly as in the former
references, we discern different driving styles, but we focus in doing this on
the energy efficiency. Then, our computational approach resembles more the
approaches taken in the latter references. Our analysis is based on a set of GPS
logs collected beforehand during driving. The processing of the data is done in
an offline post-processing step. Thus, one of our contributions is to carry over
some ideas from traffic safety engineering and related areas to a novel field of
application. Furthermore, we propose some dedicated techniques in order to meet
the requirements of our setting and provide a reliable solution in discerning
energy efficient driving styles.\par
Our ultimate goal is to classify the drivers solely based on positional
information as this is a relevant setting for potential industrial applications
of our approach. Let us stress that the use of GPS logs alone is therefore an
important issue, and this also distinguishes our work from others that mark the
current state-of-the-art in the technically related above-mentioned literature.
Despite a high amount of noise in the existing real-world data set
\cite{YZZX2010} that we employ for demonstrating our method we still obtain
robust results. Our clustering algorithm employs an improved formulation of the
jerk quantity introduced in \cite{MMK2009} and used there for driver's
behavioural analysis for traffic safety purposes.\par
Let us also stress that the sole use of positional information implies that it
does not make much sense to employ a variety of features based on these data as
these will all be naturally correlated. Having just the GPS data at hand it is
thus of primal interest to identify \emph{one} reliable and robust feature to
work with. Moreover, this feature should be meaningful by relating to the energy
efficiency of driving styles. It is exactly one of our contributions to provide
this by means of a novel jerk-based feature.\par
In combination with an agglomerative hierarchical clustering algorithm we
achieve a reliable classification result with reasonable computational effort.
Furthermore, we suggest a simple strategy to determine a reasonable amount of
clusters. Experimental results show that our approach is more robust than a
straight forward adaptation of previous approaches.\par{}
In the paper we proceed as follows. First we give a detailed physical-based
argumentation that motivates the use of the jerk as a meaningful feature for our
application. This is followed by a presentation of our clustering method for
classifying drivers. In the final section we discuss at hand of the application
of our method at the real-world data set \cite{YZZX2010} the viability of our
approach. The paper is finished by a summary and conclusion.
%
\section{Motivation: On energy efficiency and jerk}
\label{sec:2-a}
%
Let us now motivate our use of the jerk as the underlying feature of our
investigation. Thereby we comment on two important aspects in pattern
recognition, physical significance of the feature and useful invariances for our
application.
%
\subsection{Basic physical considerations}
%
For modelling we consider the movement of a car during a fixed time frame which
we parametrise via the time $t \in [0,T]$. For simplicity of notation, we assume
that the car performs a 1-D movement along a straight horizontal path $S$ in
this time frame. We denote by $s_0$ and $s_T$ the starting point and the end
point of $S$, respectively, and we write $d$ for the length of the path.
Concerning the time-dependent position of the car along $S$, we make use of a
corresponding function $x(t)$. We sometimes denote by
$v= \dot x\coloneqq\dot x(t)= v(t)$ the velocity and by $a= \ddot x = \dot v$
the acceleration of the car. We assume for simplicity that the mass $m_c$ of the
car is a constant over the considered time frame (neglecting e.g.\ mass loss due
to consumed fuel).
Another fundamental yet not too obvious assumption for our modelling is that the
driver aims to travel the distance $d$ of path $S$ during the considered time
frame over the time interval $[0,T]$. This underlying assumption is required as
otherwise one may realise a fuel saving driving style by performing a full break
and shutting off the car. Let us note that this underlying assumption is in
accordance to the content of given data after our preprocessing as GPS logs of
cars standing still will be discarded.\par
Let us now assert that in the engine, fuel is transformed into mechanical work
$W$. In order to obtain a measure for fuel consumption (denoted by
$\mathsf{fc}$), the transformation process has to be related to consumption over
the considered time frame, or equivalently to the distance $d$ passed during the
time frame. To this end we opt to measure $\mathsf{fc}$ via the generated
mechanical work over distance $d$ and assume the proportionality law:
\begin{equation}
      \mathsf{fc} \sim \frac{W}{d}
      \quad\text{or}\quad
      \mathsf{fc} = p\frac{W}{d}
      \label{hybb-add-1}
\end{equation}
where the proportionality constant $p$ is given by the combustion efficiency of
the engine.\par
Let us note that in terms of physical measurement units, we have by $[d]=m$
\begin{equation}
      [\mathsf{fc}] = \left[\frac{W}{d}\right]
      = \si{\kilogram \, \metre \per \second^{2}} = \si{\newton}
      \label{hybb-add-2}
\end{equation}
that means the fuel consumption is proportional to the force $F$ with
$[F]=\si{\newton}$ needed to move the car along $S$.\par{}
By considering now Newton's second law we have for the 1-D movement of the car:
\begin{equation}
      W = F \cdot d
      \quad \Leftrightarrow\quad
      \frac{W}{d} = F = m_c \cdot \bar a
      \label{hybb-add-3}
\end{equation}
where $\bar a$ is the acceleration needed over $[0,T]$ in order to move the car
from $s_0$ to $s_T$ over distance $d$ along $S$. As a consequence of
\eqref{hybb-add-1} to \eqref{hybb-add-3}, we obtain
\begin{equation}
      \mathsf{fc} \sim \bar{a}
      \quad\text{or}\quad
      \mathsf{fc}= p \cdot m_c \cdot \bar{a}
      \label{hybb-add-4}
\end{equation}
Let us now consider the situation that the car enters $S$ with a certain
velocity $\bar v>0$. Then by basic physical principles the kinetic energy
$E=\frac{1}{2}\si{\metre} \bar v^{2}$ corresponds to the work $W$ stored in the
movement.\par{}
In an ideal, frictionless environment the car moves on with constant speed
$\bar v$, and there is no need to install an additional acceleration to hold
$\bar v$ as would be desired by a driver in order to travel along $S$ during the
considered time frame. It is clear that in reality where e.g.\ friction takes
place, a certain acceleration is needed to allow the driver to travel the
distance $d$ along $S$ during $[0,T]$. Since $\mathsf{fc} \sim \bar a$ this also
means that a certain amount of fuel has to be used up.\par{}
The question arises how such an acceleration should be applied over $[0,T]$ in
the most fuel saving manner. One may imagine here e.g.\ that it could be
beneficial to accelerate very strongly at the beginning over $[0,\epsilon]$,
$0<\epsilon \ll 1$, and letting the car roll over the remaining time frame
$[\epsilon,T]$ to arrive then at time $T$ at point $s_T$. Let us note in this
context that the acceleration $\bar a$ just gives a total value over $[0,T]$ and
does not reveal how this total value has to be realised.\par{}
In order to identify the most fuel saving way to accelerate, it is obvious by
$\mathsf{fc} \sim \bar a$ that we have to minimise the total required
acceleration $\bar a$. To this end we now consider the main forces acting on the
car. We propose to constitute as main sources of fuel consumption
\emph{frictional forces} and \emph{aerodynamic resistance}. Since forces are
acting (by fundamental physical principles) in an additive and independent way
on the car, it will turn out that we may discuss them separately.\par{}
Before doing that, we first have another look at the basic mechanism behind fuel
consumption itself. We consider again \eqref{hybb-add-1} and formulate the fuel
consumption using the total transformed fuel over the considered time frame. For
the computation we make explicit by taking the absolute of $W$, that braking
does not generate fuel:
\begin{equation}
      \mathsf{fc} \sim \int_0^T \left| W(t) \right| \, dt 
      \stackrel{W = F \cdot d}{\sim}
      \int_0^T \left| a(t) \right| \, dt 
      =
      \int_0^T \left| \ddot x(t) \right| \, dt 
      \label{hybb-add-5}
\end{equation}
To minimise fuel consumption during the transformation process to mechanical
work, it is therefore optimal to uphold a constant velocity since the minimiser
of the above expression is obtained for
$(\ddot x = 0) \Leftrightarrow (v=\text{constant})$.\par{}
Let us turn to frictional forces $F_{friction}=\eta \cdot F_n$, where $\eta$ is
the friction coefficient and $F_n=m_c g$ the normal force described by the
constant mass of the car $m_c$ and the gravitational constant $g$. In this
context let us note that we impose indirectly that the car moves approximately
horizontally. Assuming that $\eta$ is constant over our time frame, this means
that to negate frictional forces implies to uphold a constant acceleration $a$,
compare \eqref{hybb-add-3}.\par{}
Turning to the aerodynamic resistance of a car, this may be modelled by a force
$F_{air} \sim v^2= \dot x^2$. In order to minimise the required acceleration
negating aerodynamic resistance, we thus have to find the minimiser of
\begin{equation}
      \min_x
      \int_0^T \left( \dot x(t) \right)^2 \, dt 
      \label{hybb-add-6}
\end{equation}
subject to boundary conditions $x(0)=s_0$ and $x(T)=s_T$. The corresponding
optimality condition reads as $2\dot v=0$. Therefore it is optimal w.r.t.\ fuel
consumption to uphold a constant velocity in order to negate the aerodynamic
resistance. This can be realised by a constant acceleration added to the one we
found to be required to negate frictional forces.\par{}
As a consequence of our investigation, it is optimal to negate the
fuel-consuming forces by a constant acceleration, thereby upholding a constant
velocity of the car. Since a constant acceleration implies $\dot a=0$, one may
detect instances of potential fuel-wasting driving by evaluating
$\dot a= \ddot v$, and it appears to be an energy-efficient driving style if
$\left| \ddot v \right|$ is kept low by a driver. This result is in accordance
with the intuition that drivers aiming at an energy saving driving style usually
perform smooth accelerations and braking, whereas racy, highly energy consuming
drivers tend to have a more abrupt driving style. Also dense urban traffic with
the typical changes in acceleration and deceleration which is notorious for
leading a high fuel consumption is represented by strong instances of
$\left| \ddot v \right|$. \par
%
\subsection{Invariances}
%
As discussed, the quantity $\ddot v=\dot a$ describes the variation of the
acceleration of a car. However, the positional GPS data gathered in our database
also allows us to formulate quantities that offer certain \emph{invariances}.
Our analysis should not depend on an absolute positioning and yield the same
findings whether we analyse cars in Europe or in Asia. Such an invariance can be
introduced by taking derivatives of the movement. The velocity is independent of
the exact location of a car. Further, the acceleration of a car driving smoothly
with $\SI{130}{\kilo\metre\per\hour}$ is similar to a car driving with
$\SI{50}{\kilo\metre\per\hour}$. If the environmental circumstances are
adequate, then both drivers should have the same classification.\par
This observation motivates the usage of a feature with a sufficient large set of
invariances. The quantity $\ddot v$ is invariant under affine transformations of
the velocity and offers these benefits, too.\par
%
\subsection{Formalisation}
%
We conclude our motivation by observing that the quantity $\ddot v$ is
well-known in physics as jerk. For convenience, we formalise this observation at
hand of the following definition.
\begin{definition}[Jerk function]
      The \emph{jerk} $j(t)$ describes the third order derivative of the
      position $x(t)$ with respect to time:
      $j(t) \coloneqq \frac{\mathrm{d}^{3}}{\mathrm{d} t^{3}} x\left( t
      \right)$.
\end{definition}
Let us note that we refer here to jerk in terms of the derivative of a position
$x$ since our input data is given by positions.

We also remark that in \cite{MMK2009}, the authors use already the jerk function
to classify drivers with respect to their aggressiveness, however, as we have
shown, the jerk is also a reasonable physical quantity related to energy
consumption during driving.
\begin{figure*}                                                     
  \centering
  \includegraphics[scale=0.7]{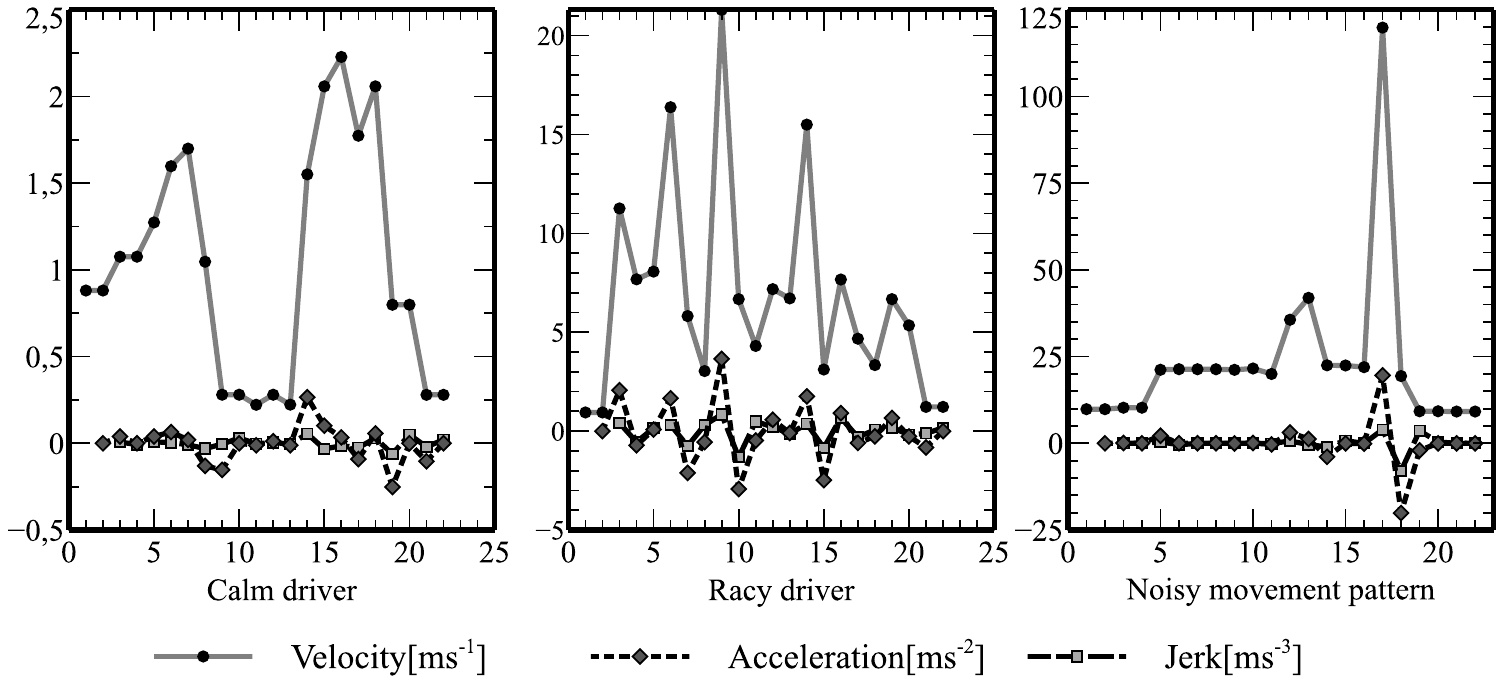}
  \caption{Velocity, acceleration and jerk patterns corresponding to a calm and
    a racy driver in contrast to the same quantities of a movement pattern
    containing noisy GPS logs. Note that the $y$-axis in each plot has a
    completely different range. Either of the acceleration or jerk pattern could
    distinguish the calm from the aggressive driver, efficiently. In case of the
    noisy movement pattern, still either of the acceleration or jerk pattern
    could efficiently reveal the existence of a noisy GPS log. Our proposed
    exponential based feature $\omega$, cf.\ \eqref{eq:feature_main}, shows high
    robustness w.r.t.\ noise. It helps to accommodate noisy movement patterns
    totally in a separate cluster.}
  \label{fig:vel_acc_jrk}
\end{figure*}
%
\section{Driving style analysis by GPS data}
\label{sec:2}
%
Relying exclusively on GPS data, as provided by navigation systems and tracking
devices, comes with a certain number of hurdles that need to be overcome. Our
goal is to classify drivers with respect to their driving style. Currently, this
analysis is done offline. We first collect the data and store it in a database.
The processing and classification is done afterwards. To this end we measure the
drivers' spatio-temporal positions.\par
Definition~\ref{thm:1} below gives us a formal framework to work in. Positional
data of the cars comes in form of coordinate pairs
$\left(x_{i},y_{i}\right)\in\mathbb{R}^{2}$. Each pair is accompanied with a
label containing the time stamp $t_{i}$ when the position has been recorded.
This set of discrete samples gives us a complete description of the movement of
a car in space and time. Yet, our focus lies on the analysis of moving cars.
Therefore, we use a more specific structure to represent the displacement of a
vehicle.
\begin{definition}[Time frame, Movement pattern]
  \label{thm:1}
  We define a \emph{time frame of size $\ell$} a sorted vector
  $T\in\mathbb{R}^{\ell}$ containing $\ell$ individual time stamps in
  non-decreasing order. We call \emph{movement pattern of size $\ell$} a pair
  $\{T,P\}$ consisting of a time frame $T$ of size $\ell$ with the corresponding
  positional data
  $P\coloneqq\{\left(x_{i},y_{i}\right)\vert\; i=1,\ldots,\ell\}$ if the car is
  not standing still during any moment within the complete time frame.
\end{definition}
Thus, we speak of a movement pattern if the car does not stop during a
considered time interval. We elaborate a method to discard the idling objects
from our data set in the next section.
%
\subsection{Data preprocessing}
\label{subsec:21}
%
In order to apply our method we need to preprocess the given data set of GPS
logs. First of all, we discard corrupted or meaningless data. After the
preprocessing our data should only consist of movement patterns as defined in
Definition~\ref{thm:1}. To this end we remove all logs for which the following
conditions hold:
\begin{enumerate}
  \item All GPS logs having the same longitude and latitude values as well as
        the same sampled time as their adjacent logs. These logs represent
        repeated GPS values being sampled at least twice due to hardware
        failure.
  \item All GPS logs with same longitude and latitude having different sampled 
        time compared to their adjacent GPS logs. These GPS logs represent a car
        standing still.
\end{enumerate}
The preprocessing is necessary to ensure reasonable findings. However, it also
introduces holes in the displacement history of the analysed car. Certain
records are contiguous over long time intervals whereas others might only run
for a few seconds before a gap occurs. Our forthcoming analysis is based on
statistical measures of the movement pattern. To allow a meaningful and unbiased
comparison we additionally drop movement patterns that are too short and break
down those that are too long into several smaller ones. In this work we discard
all movements patterns with less than 10 samples and split them if they exceed a
length of 24. After the preprocessing we obtain about 3225 movement patterns
from the real-world data set \cite{YZZX2010}.\par{}
%
\subsection{Clustering movement patterns by a novel jerk based feature}
\label{subsec:22}
%
Let us implement our underlying considerations about the jerk in the context of
a typical example of GPS data.
\begin{figure*}                                                                             
  \includegraphics[scale=0.9]{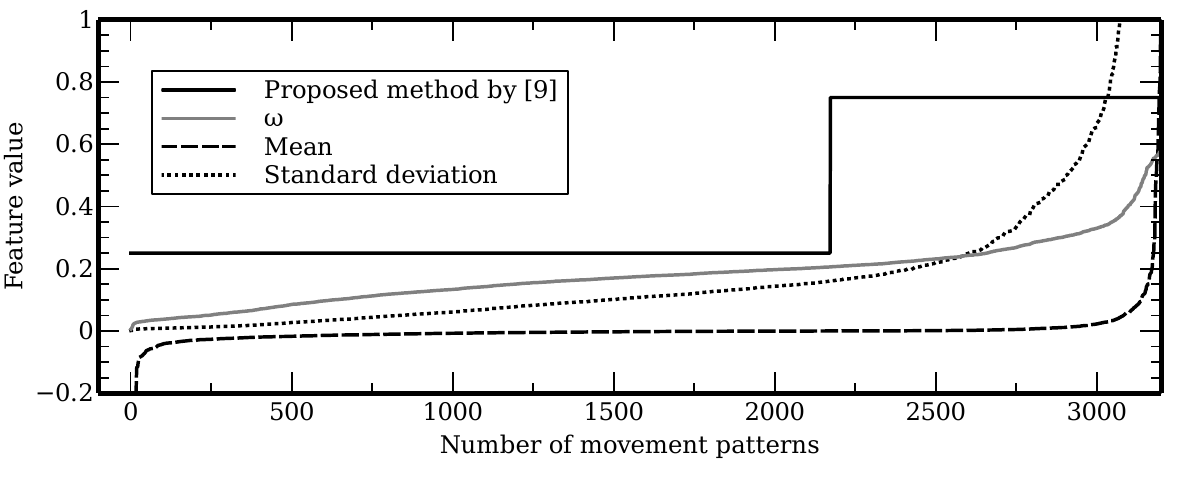}                                 
  \caption{The sorted values of mean, standard deviation, the feature proposed
    by \cite{MMK2009} and our proposed feature for all preprocessed movement
    patterns from \cite{YZZX2010}. The approach from \cite{MMK2009} only yields
    two distinct classes and cannot detect noise. On the other hand, the curve
    representing the mean is almost flat, rendering it very difficult to
    distinguish different clusters. The standard deviation depicts a similar
    behaviour as our proposed feature but bears a less stable behaviour towards
    the end. Our feature yields a curve with a clear monotonic increase and a
    notable elbow at the end. Thus, we obtain the same segregation ability as
    the standard deviation but also a clearer threshold for the noise detection,
    which is given by the samples located in the elbow.}
  \label{fig:omegafeature}
\end{figure*}
As discussed, important physical quantities in the description of the behaviour
of drivers are the position, velocity, and acceleration of their car.
Figure~\ref{fig:vel_acc_jrk} depicts an example of velocity, acceleration, and
jerk obtained from a movement pattern of a calm driver, a racy driver, along
with a movement pattern containing a noisy log. Even though a distinction among
the three shown patterns could be deduced from the velocity and acceleration
alone, the jerk has the benefit of magnifying big abrupt changes in the
acceleration much more than smaller ones and therefore, simplifies the noise
detection significantly. This can clearly be observed in the noisy movement
pattern shown in Fig.\ \ref{fig:vel_acc_jrk}. The previous arguments together
with our physical considerations constitute the principal motivations to use the
jerk as starting point in our work. Our exponential based feature proposed in
\eqref{eq:feature_main} takes advantage of the elaborated properties.\par{}
The authors of \cite{MMK2009} use the ratio between the standard deviation of
the jerk function and its mean value as feature to distinguish between calm,
normal and aggressive drivers. Their method uses a data set obtained from
\cite{Sierra2009}. The movement patterns are augmented with additional
information that indicates the road type (e.g.\ freeway, ramps, local roads,
etc.) as well as the mean jerk value expected for the corresponding road type.
This information is usually not available in standard GPS logs. Furthermore, the
classification from \cite{MMK2009} uses a hard threshold. We believe, that a
soft thresholding is better suited for classifying such complex patterns as
driving styles. To tackle the just mentioned problems we propose a few adaptions
in the forthcoming paragraphs.\par{}
%
\paragraph{Modelling details}
%
Our proposed method seeks to classify driving style using an alternative
formulation which also considers the jerk. We suggest to use a clustering
strategy exclusively based on the following feature:
\begin{equation}
  \label{eq:feature_main}
  \omega\left( {\{T, P\}} \right) \coloneqq
  \exp\left( -\frac{1}{\sqrt{\sigma\left(j_{\{T, P\}}\right)}}\right)
\end{equation}
Here, $j_{\{T, P\}}$ is the array of jerk values for each position in the
movement pattern $\{T, P\}$. These are computed by means of standard finite
difference schemes.\par
The function $\sigma$ denotes the standard deviation. The motivation for
proposing this feature is as follows. The exponential part of
\eqref{eq:feature_main} discriminates those movement patterns having a very high
amount of fluctuation rate around their mean jerk values. They occur due to the
existence of at least one noisy GPS log inside the movement pattern or a dubious
driving style with a large number of strong accelerations and decelerations.
Further, in order to discern drivers with lower jerk fluctuation rates we
consider the square root of the standard deviation inside the exponential
function. This facilitates the formation of a smooth elbow in our proposed
feature $\omega$, see Fig.~\ref{fig:omegafeature}.\par
In our experiments, the modified jerk feature yielded the most intuitive and
reliable clustering results. Small feature values correspond to those movement
patterns representing drivers with less accelerations and decelerations in their
driving patterns. An energy-saving driving style can therefore be identified by
small feature values, whereas more racy drivers will usually exhibit larger
feature values.\par{}
%
\paragraph{Algorithmic details}
%
Two classic and well studied approaches to classify data are partitional and
hierarchical clustering methods. The latter have the advantage to yield a
complete scale evolution for all possible numbers of clusters. Since the optimal
number of clusters is hard to predict, we opt for such a flexible approach using
an agglomerative clustering method. Such a method starts by taking singleton
clusters at the bottom level and continues merging two clusters at a time to
build a bottom-up hierarchy of clusters.\par
We employ Ward's criterion for the merging strategy \cite{W1963,W1969}. It makes
use of the standard k-means squared error (SSE) to determine the distance
between two clusters. For any two clusters, $C_{a}$ and $C_{b}$, Ward’s
criterion is calculated by measuring the increase in the value of the SSE for
the clustering obtained by merging them into a single cluster
$C_{a}\cup{}C_{b}$. The characteristic number of Ward’s criterion is defined as
follows:
\begin{equation}
  \label{eq:2}
  \frac{\abs{C_{a}}\abs{C_{b}}}{\abs{C_{a}}+\abs{C_{b}}}
  \sum_{\nu=1}^{\ell}
  \sum_{\eta=1}^k
  \left( {c_{a}}_{\nu} - {c_{b}}_{\eta}\right)^{2}
\end{equation}
where the cluster $C_{x}$ has the individual components ${c_{x}}_{\nu}$. The
cardinality of a cluster is denoted by $\abs{C_{x}}$.\par
We iteratively merge clusters in a bottom-up fashion. At the bottom level, each
computed feature point is considered. Then, at each new level we merge the pair
of clusters that minimises \eqref{eq:2}. The algorithmic details of the
agglomerative hierarchical clustering are also given in
Algorithm~\ref{alg:alg-clust}.\par{}
\begin{algorithm}[tb]
  \DontPrintSemicolon
  \KwData{A set of movement patterns.}
  \KwResult{A clustering of the input movement patterns.}
  Compute the feature value $\omega$ from \eqref{eq:feature_main} for each
  movement pattern.\;
  Place each feature value in its own cluster.\;
  \Repeat{a single cluster is left}{
    Find pair of clusters that minimises \eqref{eq:2}.\;
    Merge this pair into a single cluster.\;
  }
  \Return the smallest number of clusters, that yielded a decrease in
  \eqref{eq:3} below a threshold $T$.\;
  \caption{Agglomerative hierarchical clustering of movement patterns}
  \label{alg:alg-clust}
\end{algorithm}
The optimal number of clusters is obtained through an L-curve heuristic
\cite{H1992,HO1993}. We compare the number of clusters against the
within-cluster sum of squares (WCSS) and consider the smallest number of
clusters that yields a decrease in the WCSS below a given thresh\-old. The WCSS
is given by:
\begin{equation}
  \label{eq:3}
  \sum_{a=1}^{M}\sum_{x\in C_{a}} \norm{x-c_{a}}_{2}^{2}\enspace{}
\end{equation}
where $M$ is the total number of clusters and $c_{a}$ is the centroid of cluster
$C_{a}$.\par
\begin{figure*}
      \hfill
      \subfigure[Here the number of clusters is considered to be three. In this case a group of 
      drivers in the $2^{nd}$ cluster is probably not optimally labelled. It is 
      not possible to decide if they belong to average fuel consuming drivers 
      or to noisy patterns. Only the energy saving drivers could be clearly located in 
      the $1^{st}$ cluster.]
      {\includegraphics[width=0.32\textwidth]{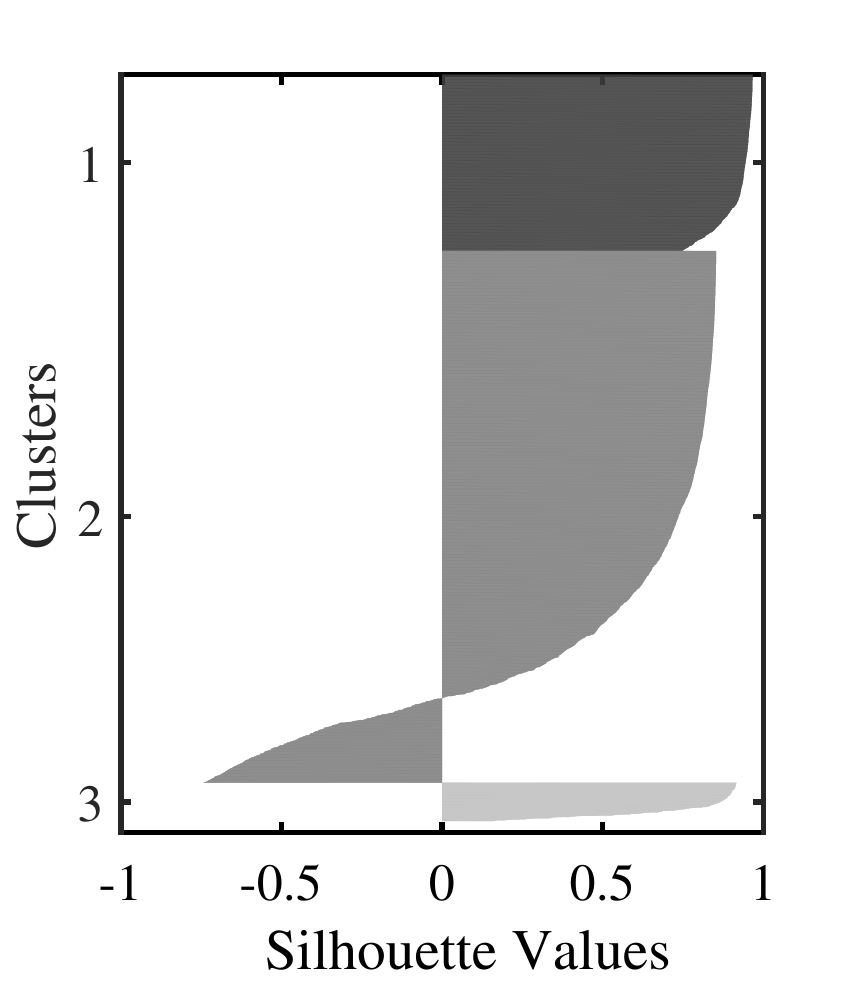}}
      \hfill
      \subfigure[A minimal number of drivers with a negative silhouette index is
      found with 4 clusters. This clustering seems to be the best labelling. Noisy
      driver patterns are likely to be found in the $4^{th}$ cluster. All energy
      saving drivers are located in the $1^{st}$ cluster and average fuel consuming
      drivers expand across two classes in the $2^{nd}$ and $3^{rd}$ cluster,
      respectively.]
      {\includegraphics[width=0.32\textwidth]{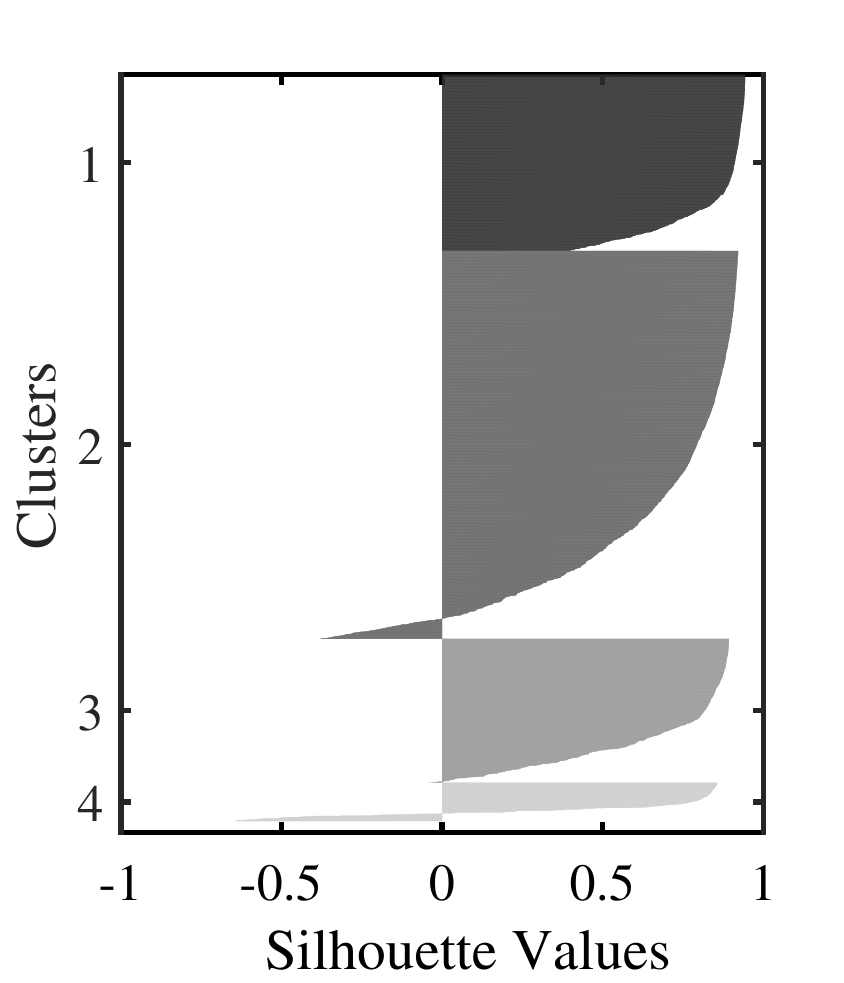}}
      \hfill
      \subfigure[By increasing the number of clusters to five, drivers with 
      negative silhouette indices appear even in the $1^{st}$ cluster with small 
      $\omega$ values. This depicts a clustering result which may not be considered 
      as optimal, since almost all clusters accommodate dissimilar drivers inside them.]
      {\includegraphics[width=0.32\textwidth]{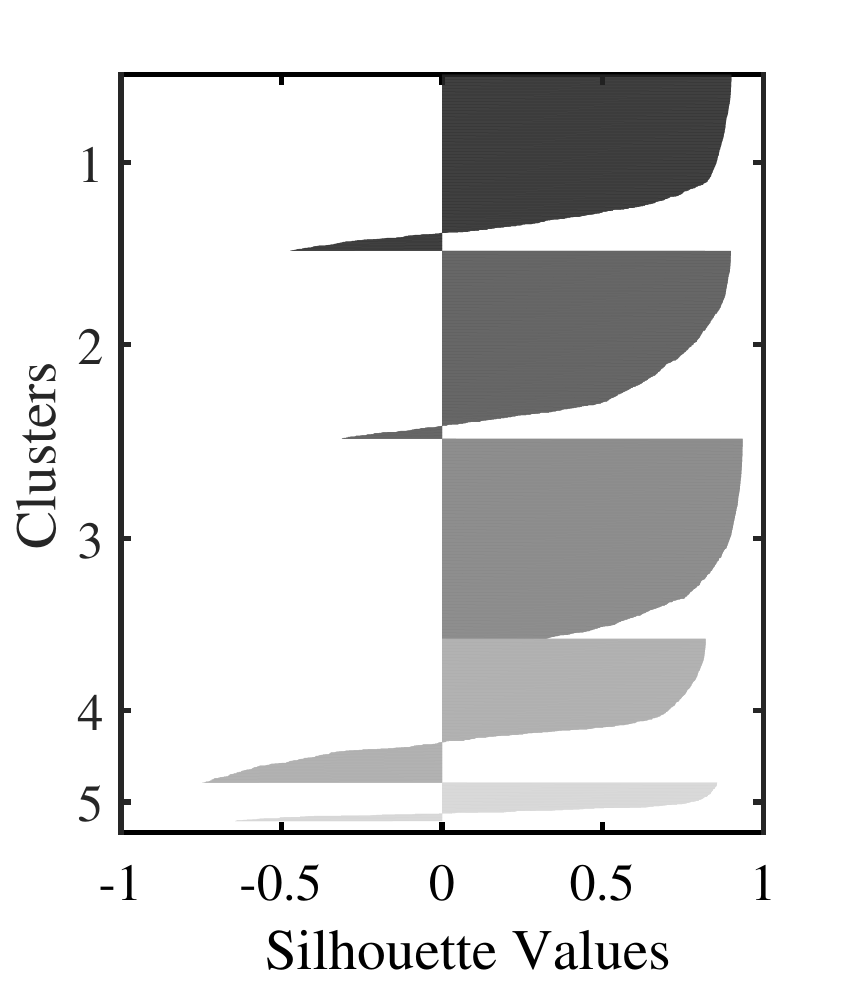}}
      \hfill
      \caption{The silhouette indices of a clustering setup establishing (a) three, 
        (b) four and (c) five distinct groups of $\omega$ values of the central area 
        drivers in Beijing. Inside each cluster, the more positive a silhouette index of
        a feature, the more similar is the feature to its group-mates which 
        indicates a good clustering.}
      \label{fig:Sill_4_cluster}
\end{figure*}
In our WCSS based experiments the optimal number of clusters was usually found
to be four. In what follows, the silhouette index $SI_{i}$ \eqref{eq:sill_index}
is computed for members of each four established clusters:
\begin{equation}
  \label{eq:sill_index}
   SI_{i} = \frac{b_{i}-a_{i}}{\max(a_{i},b_{i})} 
\end{equation}
Here, $a_{i}$ is the mean distance between any cluster member $\omega_{i}$ to
all other members in same cluster and $b_{i}$ is the mean distance among the
cluster member $\omega_{i}$ and all other members in next nearest cluster. The
computed $SI_{i}$ corresponding to $\omega_{i}$ takes negative or positive
values in range of $[-1,+1]$ as being mostly dissimilar or similar to other
members in same cluster. Well established clusters require silhouette indices to
be positive over their members.

Our best clustering result is analysed with respect to its labelling in
Fig.~\ref{fig:Sill_4_cluster}~(b). The negative silhouette indices inside the
$4^{th}$ cluster represent the compactly accommodated noisy $\omega$ values.
This observation validates our approach for identifying the noisy values. In
addition, only a few members in the $2^{nd}$ cluster are wrongly located making
the final clustering result meaningful.\par
Let us emphasise that an alternative strategy would also be possible with the
k-means algorithm. However, k-means needs to be restarted for each new amount of
clusters and thus adds a significant computational burden if the number of
clusters is not predetermined exactly.\par{}
%
\section{Experimental results and validation}
\label{sec:3}
%
In order to investigate the efficiency of our proposed feature we consider a
database with GPS logs from two distinct parts of Beijing city \cite{YZZX2010}.
The whole database contains the trajectories of more than 10,000 taxis collected
during the period of Feb.\ 2 to Feb.\ 8, 2008. In total there are more than 17
million GPS logs that cover a total distance of around 9 million kilometres.
Since the data set corresponds to various taxi drivers we expect to find many
different driving styles in this data set. The regions that we consider contain
a vast area of Beijing city centre and a freeway expanded from east to west. We
preprocess the logs as described in Section~\ref{subsec:21} and obtain
correspondingly 3225 patterns for the city centre and 106 movement patterns for
the freeway. All our feature values are computed according to
\eqref{eq:feature_main}.\par{}
In a first experiment we compare the Beijing city centre feature values as by
\eqref{eq:feature_main} to other commonly used feature choices. The findings are
visualised in Fig.~\ref{fig:omegafeature}. We compare our suggested feature
against the mean value and standard deviation as well as the jerk-based feature
from \cite{MMK2009}. The approach from \cite{MMK2009} (proposed for the purpose
of analysing driver behaviour, not for energy efficiency) yields a clear cut. It
segregates our drivers into two categories, namely defensive and aggressive
drivers as by the context of the work \cite{MMK2009}. For this classification we
used the parameter suggestions from \cite{MMK2009}. We selected a window size of
10 seconds and set the parameters norm\textsubscript{threshold} and
agg\textsubscript{threshold} to 0.5 and 1 respectively. This method yields a
very clear threshold that discerns the two classes of drivers. However, it is
unable to detect noisy data. If we consider the average value and standard
deviation as features then we are able to detect the corrupted data but a
discrimination of the remaining drivers becomes difficult. All their feature
values are very close to each other. Finally, our proposed feature yields a
larger range of values as well as a clear jump at the end of the curve. Thus, we
are in the position to distinguish different driving styles as well as filtering
out noisy patterns.\par
\begin{figure*}
      \centering
      \includegraphics[scale=0.55]{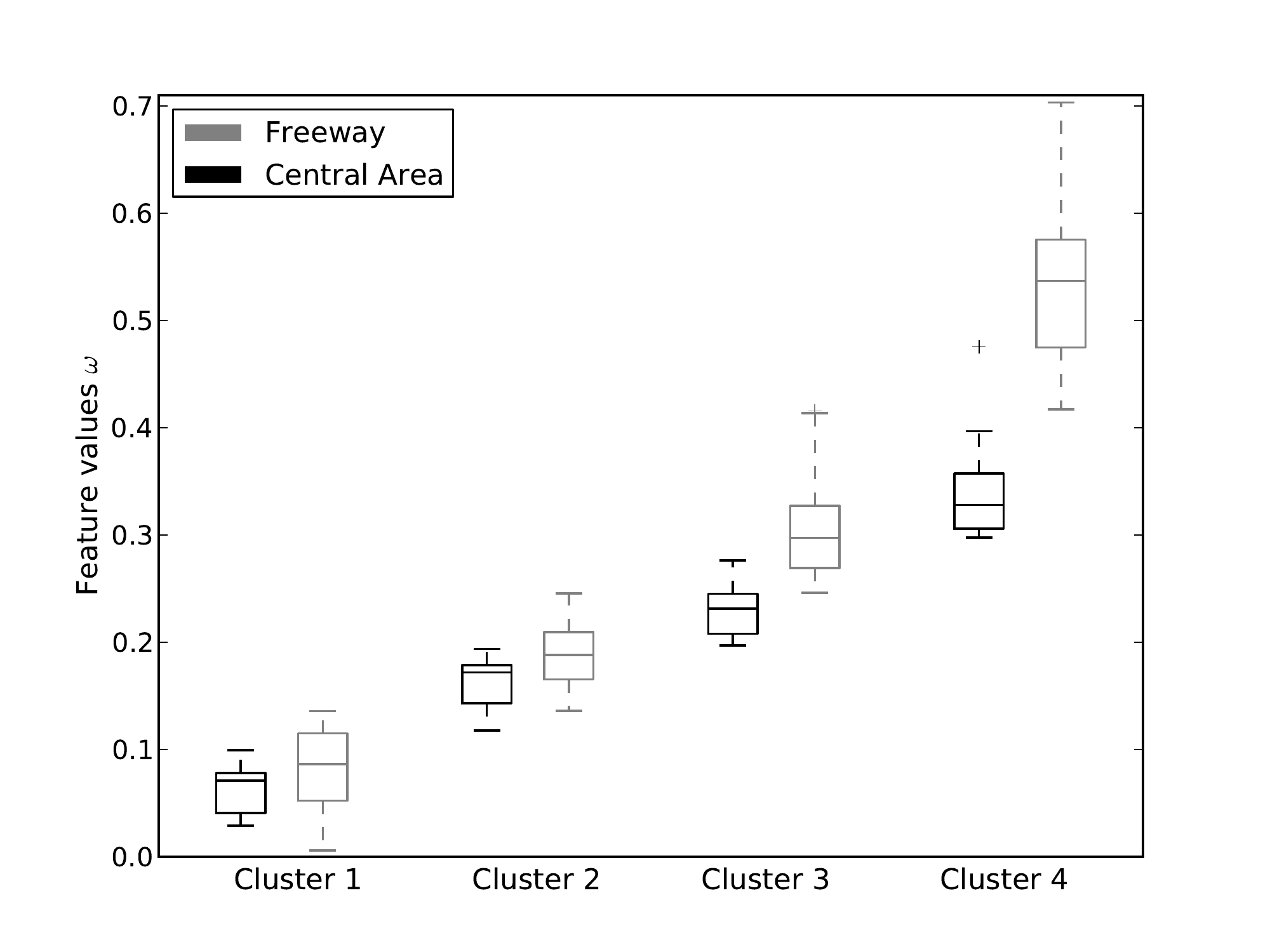}
      \caption{A whisker plot showing the distribution of the feature values
        among the different clusters for our two experimental setups. The box
        marks the boundaries of the first and third quartiles, while the bars
        indicate the full extent of the feature values in that cluster. The
        horizontal line indicates the median. As we can see, the first cluster
        and on the other hand the second and third cluster are well separated in
        terms of feature values. There is a continuous overlap from the second
        to the third cluster. These two clusters represent the majority of the
        drivers with an average fuel consuming driving style, with tendency
        towards being more (cluster 2) and less (cluster 3) energy efficient.
        The first cluster represents the energy saving drivers and the last
        cluster the drivers with a high fuel consumption, containing also noisy
        driver patterns.}
      \label{fig:ClusterDist}
\end{figure*}

Next, we evaluate our proposed approach by applying the agglomerative clustering
detailed in Algorithm~\ref{alg:alg-clust}. The distribution of all feature
values among the different clusters is visualised in Fig.~\ref{fig:ClusterDist}.
As we can see, the feature values from the first cluster and the fourth cluster
are clearly different from those in the second and third cluster. It confirms
our expectations. The first cluster contains the energy saving drivers, whereas
cluster 4 is supposed to contain noise as well as the highly fuel consuming
drivers. Finally, the majority of the drivers lies within the bounds of the
second and third cluster and present an average fuel consumption.\par
%
\paragraph{Findings for Beijing city centre area}
%
The clusters 1, 2, and 3 are totally noise free and could be adopted as an
accurate driver's behavioural model. The within-cluster sum of squares (WCSS)
index does not show any remarkable amount of reduction by adding a
$5^{\text{th}}$ cluster or more. Hence, according to our L-term heuristic, we
should set the final number of clusters to 4. The noisy patterns are located
fully in cluster number 4.\par{}
Let us stress that our algorithmic proceeding that results in setting the number
of clusters to four is confirmed in an independent way by evaluating the
silhouette indices as shown in Fig.~\ref{fig:Sill_4_cluster}. This is shown here
exemplarily for the data of the Beijing city centre area but also holds for the
other analysed data.
%
\paragraph{Findings for the freeway area}
%
The clusters 1 resp.\ 2 are noise free. The L-term strategy suggests to use 4
clusters in this case. Since we only have very few feature values in this area
our algorithm does not yield a cluster solely containing noise. However, the
$4^{th}$ cluster contains partially the noisy movement patterns. Such a cluster
was present for the Beijing centre area, where the $4^{\text{th}}$ cluster was
filled with corrupted patterns only. This observation also shows that our
classification algorithm benefits from having an extremely large amount of
samples to its avail. This is a realistic scenario in an industrial application
where large amounts of data can be collected e.g.\ via navigation systems. The
more samples we can process the clearer we can distinguish noise and classify
the individual drivers.
%
\section{Summary and conclusion}
\label{sec:conclusion}
%
In this paper we have shown that it is possible to discern different driving
style patterns with respect to their energy consumption from their GPS logs
alone. To this end we use a dedicated variation of the jerk feature and combine
it with a hierarchical clustering approach.\par
Our model is quite simple but still capable of discriminating drivers into
different classes and filtering out noisy data logs. Let us note again that the
use of just the GPS logs is highly relevant for a potential industrial use of
our results in the context of navigation systems.\par{}
In the future we aim to combine the driver model information with other
optimisation tools mentioned in the introduction in order to improve the energy
efficiency of hybrid cars.\par{}
%
%
\bibliographystyle{plain}      
%
\bibliography{references}
\end{document}